\documentclass[graybox]{svmult}

\usepackage{mathptmx}       
\usepackage{helvet}         
\usepackage{courier}        
\usepackage{type1cm}        

\usepackage{makeidx}         
\usepackage{graphicx}        
\usepackage{multicol}        
\usepackage[bottom]{footmisc}

\makeindex             
\begin{document}
\def\Msun{{M_\odot}}
\def\Zsun{{Z_\odot}}
\def\Lsun{{L_\odot}}

\title*{Stellar archeology: a cosmological view of dSphs}
\author{Stefania Salvadori} 
\institute{Stefania Salvadori \at Kapteyn Astronomical Institute, Landleven
12, 9747 AD Groningen (NL)\\ \email{salvadori@astro.rug.nl}}
\maketitle
\abstract{The origin of dwarf spheroidal galaxies (dSphs) is 
investigated in a global cosmological context by simultaneously 
following the evolution of the Galaxy and its dwarf satellites. 
This approach enable to study the formation of dSphs in their 
proper birth environment and to reconstruct their own merging 
histories. The proposed picture simultaneously accounts for 
several dSph and Milky Way properties, including the Metallicity 
Distribution Functions of metal-poor stars. The observed features 
are interpreted in terms of physical processes acting at high redshifts.}
\section{Cosmological background}
\label{sec:1}
Today living metal-poor stars represent the fossil records of the 
early cosmic star formation. Indeed they are expected to form at 
high redshifts, as soon as the metal (and dust) content of the gas 
was high enough $Z > Z_{cr} = 10^{-4\pm 1}\Zsun$ \cite{Schneider} 
to trigger the formation of low-mass stars, which can survive until
the present days. Extremely metal-poor stars ([Fe/H]$< -3$) have been 
observed in the Milky Way (MW) halo since long time ago \cite{Ryan}. 
However, the most promising objects to host these elusive stellar 
population are expected to be dwarf spheroidal galaxies (dSphs) that 
are metal-poor and old stellar systems, {\it all} of them showing 
the presence of a stellar population $>10$~Gyr old, despite of very 
different star formation histories.

Recently, extremely metal-poor stars have been observed in nearby dSph 
galaxies \cite{Martin}. However, while [Fe/H]$<-3$ stars represents the 
$25\%$ of the total stellar mass in ultra-faint dwarfs (UFs, $L < 10^5 
\Lsun$), they are extremely rare in the more luminous ''classical'' dSph 
galaxies ($\leq 6\%$ \cite{Else}). How did ultra-faint and classical 
dSphs form? More recently, high-resolution spectroscopic studies 
\cite{Frebela,Frebelb,Koch,Norris,Martin} have shown that, despite of 
this difference in number, the abundance patterns of [Fe/H]$<-3$ stars 
is mainly consistent in these two class of galaxies. Hence, what is the 
origin of extremely metal-poor stars?

In this contribution I will present a possible {\it cosmological scenario} 
for the origin of dSphs, by investigating the formation of these galaxies 
along the build-up of the Milky Way. Other than simultaneously accounting 
for several dSphs and MW properties, the model will allow us to physically 
explain the above mentioned features pertaining to extremely metal-poor stars. 
\section{Star formation \& feedback processes at high redshifts}
\label{sec:2}
\begin{figure}[b]
  \includegraphics[height=.41\textheight]{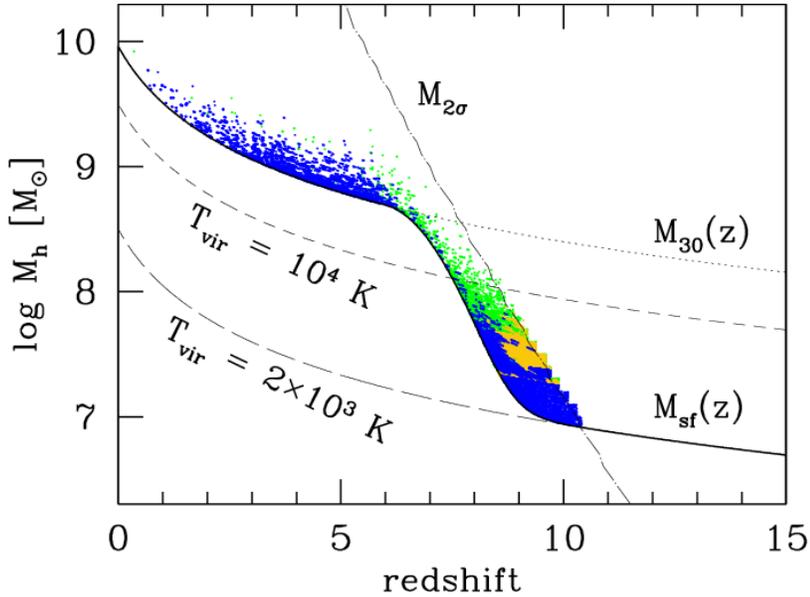}
  \caption{Hosting halo mass and circular velocity of dSph candidates (points)
    in 10 possible MW merger histories as a function of the formation redshift. 
    The gray scaling indicates the baryonic fraction $f_b$ at the formation epoch 
    with respect to the cosmic value $f_c=0.156$: $f_b/f_c > 0.5$ (dark), $0.1 < 
    f_b/f_c < 0.5$ (intermediate), $f_b/f_c < 0.1$ (light). The color version is
    available in \cite{SF09}. The lines in the panel show the evolution of
    $M_{sf}(z)$ (solid), the halo mass corresponding to 2$\sigma$ peaks (dotted-long
    dashed), $T_{vir}=10^4$K (short dashed line) and $T_{vir}=2\times 10^3$K (long
    dashed line).}
  \label{fig:M_z}
\end{figure}
The most relevant properties of the model can be summarized in few points
(for an exhaustive description see \cite{SSF07} and \cite{SF09}). 
A statistical significant sample of possible MW hierarchical merger histories 
is first reconstructed by using a Monte Carlo algorithm \cite{SSF07} based 
on the Extended Press \& Schecter theory \cite{Bond}. 
Hence, the evolution of gas and stars is followed along 
the hierarchical trees by assuming that in each halo the star formation (SF) 
rate is proportional to the mass of cold gas, whose gradual accretion 
is regulated by a numerically calibrated infall rate \cite{SFS08}.\\
Additional hypothesis are required in order to include the effects of 
{\it radiative feedback} at high redshifts, influencing both the minimum halo mass 
required to form stars, $M_{sf}(z)$, and the SF efficiency of the star forming objects. 
Indeed the gradual reionization of the MW environment, completed by $z_{rei}$, 
suppress the SF in progressively more massive haloes. The evolution of 
$M_{sf}(z)$ adopted here (Fig.1) accounts for this effect, and assume 
$z_{rei}=6$. 
On the other hand, UV photons in the Lyman-Werner band can easily 
photodissociate H$_2$ molecules, which represent the only cooling agents of 
$T_{vir}<10^4$~K ``minihaloes''. We then assume that the SF efficiency of 
these objects is reduced by a factor $[1+(T_{vir}/2\times 10^4{\rm K})^{-3}]
^{-1}$ with respect to more massive H-cooling haloes \cite{SF09}.\\
By including the effects of {\it mechanical feedback} driven by supernovae 
(SN) energy deposition (we only include the effect of SN type II \cite{SFS08}), 
we follow self-consistently the chemical evolution of the gas in both the 
proto-galaxies and in the MW environment (Fig.2) i.e. the medium out of which 
the haloes virialize and accrete gas. The SF and the SN wind efficiencies are 
calibrated to {\it simultaneously} reproduce \cite{SSF07} the global properties 
of the MW (stellar/gas mass and metallicity) and the Metallicity Distribution 
Function of Galactic halo stars \cite{Schoerck}.
\section{The birth environment}
\label{sec:3}
\begin{figure}[ht]
  \includegraphics[height=.41\textheight]{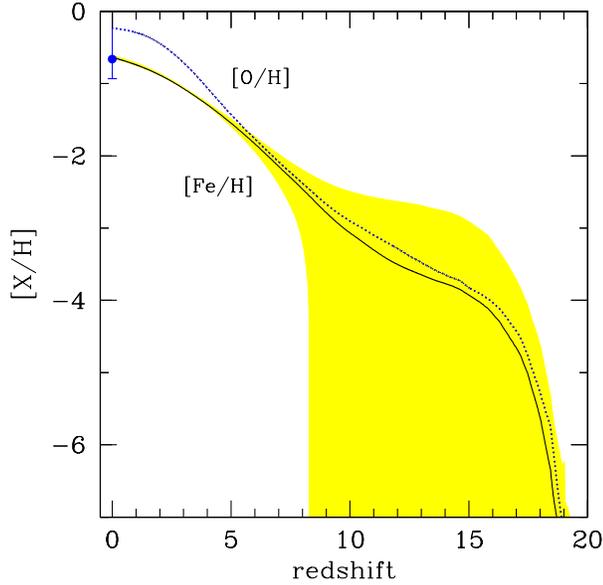}
  \caption{Evolution of the iron (solid line) and oxygen (dotted line) abundance 
  in the Galactic environment. Lines are the average values over 100 possible MW 
  hierarchical merging histories. The shaded area delimits the $\pm 1\sigma$ 
  dispersion region for [Fe/H]. The point is the measured [O/H] in high velocity 
  clouds \cite{Ganguly}.}
  \label{fig:GM} 
\end{figure}
DSph candidates are selected among the star forming haloes ($M>M_{sf}(z)$)
which are likely to become satellites i.e. those corresponding to density 
fluctuations $<2\sigma$ \cite{Diemand}. Their total dark matter mass, 
formation redshift and initial baryonic fraction $(f_b$) with respect to 
the cosmic value $f_c=\Omega_b/\Omega_m=0.156$, are shown in Fig.~1 (see
\cite{SF09} for more details). 
Note that, according to this picture, the hosting haloes of dSph galaxies 
are expected to form when $z<10.5$. In Fig.~2 the metallicity evolution 
of the MW environment is shown. We can note that the scatter is very big at 
high redshifts, reflecting the dispersion among different hierarchical merger 
histories and that the {\it average} iron-abundance rapidly increases at 
decreasing redshifts reaching [Fe/H]$\approx -4$ when $z=15$. This implies 
that {\it the MW satellites form out of a birth environment that is naturally 
pre-enriched} due to SN explosions along the build-up of the Galaxy \cite{SFS08}.
\section{Mass and formation epochs}
\begin{figure}[ht]
  \includegraphics[height=.41\textheight]{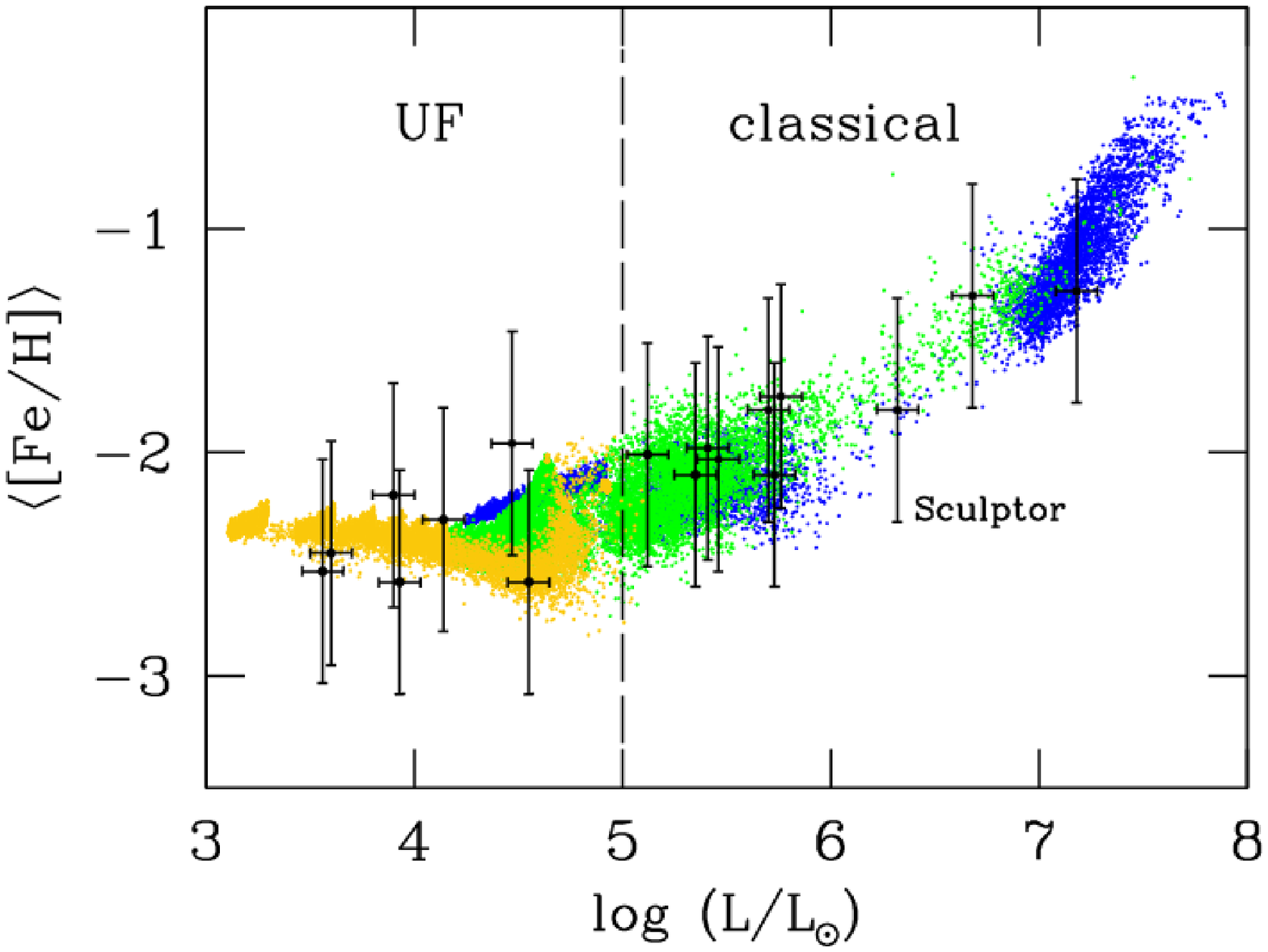}
  \caption{Total luminosity of the dSph candidates selected in 10 possible MW
    merger histories as a function of their average iron-abundance (colored points). 
    The gray scaling shows the baryonic fraction $f_b$ at the formation epoch with 
    respect to the cosmic value as in Fig.~1 (the color version is available in 
    \cite{SF09}). Points with error bars are the observational data collected by  
    \cite{Kirby}.}
  \label{fig:Fe_L}
\end{figure}
In Fig.~3 the observed and simulated Iron-Luminosity relations are compared.
The faint end of the relation, $L<10^6\Lsun$, is predominantly populated by 
$T_{vir}<10^4$~K minihaloes, in which the SF is strongly reduced due to ineffective 
molecular hydrogen cooling (Sec.~2). Above that luminosity H-cooling haloes 
dominates. We then infer that {\it ultra-faint dSphs are the living fossils 
of star forming minihaloes}. As these objects virialize when $z>7.5$ and have 
total masses $M\approx 10^{7-8} \Msun$ (Fig.~1), we conclude that {\it ultra-faint
dwarfs are the oldest and the more dark matter dominated ($M/L>100$) dSphs in the 
MW system} \cite{SF09}. On the contrary, classical dSph galaxies are associated 
to more massive H-cooling haloes, which more efficiently cool down their gas and 
assembled at later times. Sculptor-like dSphs, for example, have total masses 
$M\approx 10^{8.5} \Msun$, and finally form when $z=6.5$. Can we find the imprints 
of such a different origin in the observed properties of dSphs? 
\section{Metallicity Distribution Functions}
\begin{figure}[ht]
  \includegraphics[height=.47\textheight]{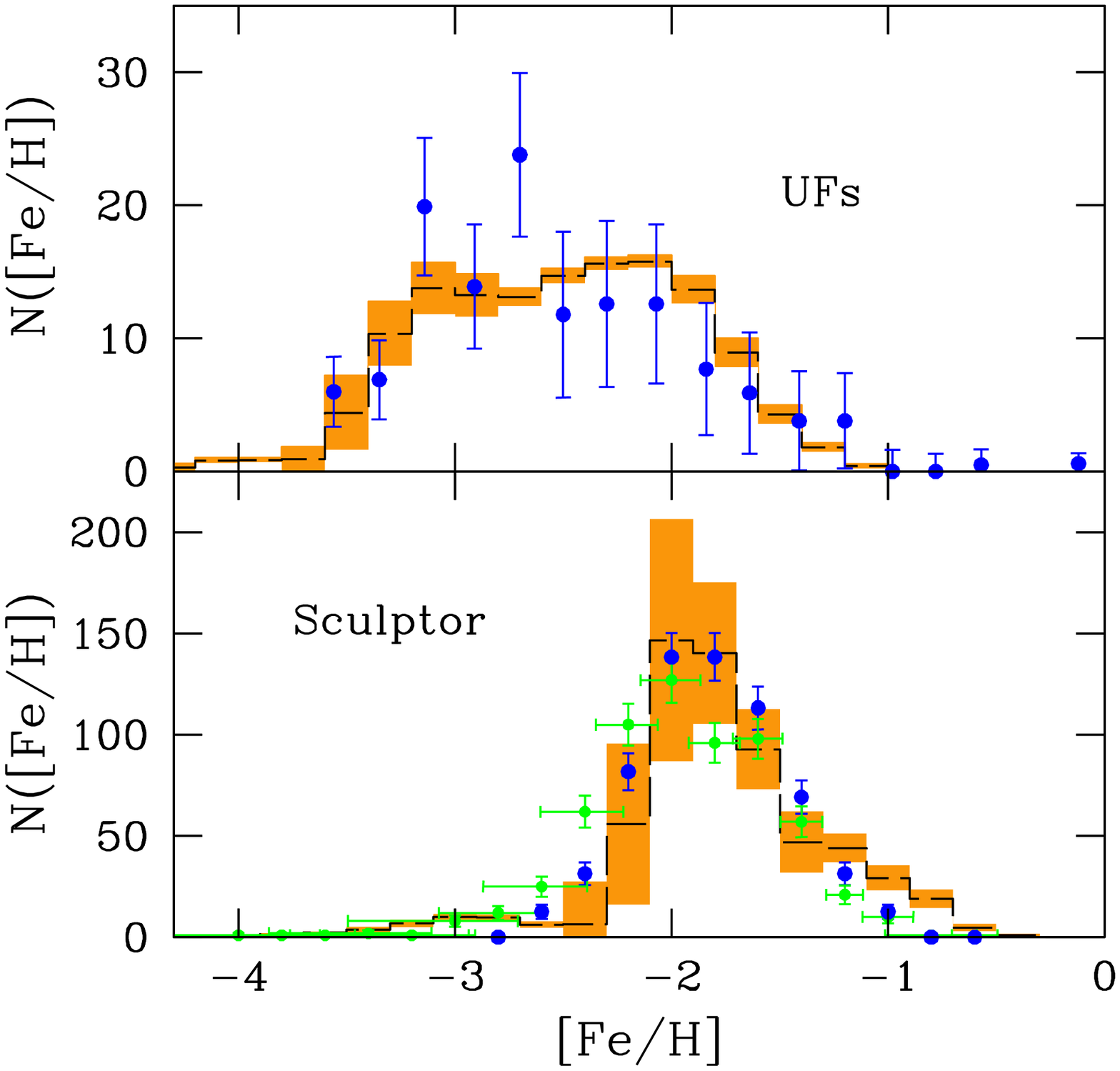}
    \caption{Observed (points) and simulated (histogram) MDF of UFs ({\it top})
    and Sculptor dSph ({\it bottom}). Histograms are the averaged MDF over all
    UFs ($L<10^5L_{\odot}$ {\it top}) and Sculptor ($10^6 L_{\odot}<L<10^{6.5} 
    L_{\odot}$ {\it bottom}) candidates in 10 merger histories. The shaded area
    is the $\pm 1\sigma$ scatter among different realizations. The data points
    are by \cite{Kirby} ({\it top}, Poissonian errors) and by the DART team
    ({\it bottom}) using the old (darkest points, Poissonian errors \cite{Helmi})
    and the new (lightest points, observational errors \cite{Else}) CaT line
    calibration.}
\label{fig:MDFs}
\end{figure}
Let's compare the average MDF of ultra-faint and Sculptor-like candidates with 
the observational data (Fig.~4). The agreement is quite good, thus allowing us 
to interpret the different MDF shapes in terms of physical mechanisms. According 
to our results the width of the MDFs reflects the cooling-efficiency of the gas: 
ultra-faint dwarfs have a broader distribution because of the extremely low SF 
rate ($<0.01 \Msun$/yr) caused by ineffective H$_2$-cooling. Few supernovae 
explode in these galaxies, which therefore slowly enrich and eject their gas. 
Sculptor-like dSphs, instead, reach high SF values ($>0.1\Msun/$ yr) at their 
final assembling epoch, thus experiencing a rapid metal enrichment along with 
a strong gas ejection \cite{SFS08}. The pronounced peak of the MDF and the 
prompt decrease for [Fe/H]$ > -1.5$ reflect these features \cite{SFS08}.\\ 
On the other hand, the shift of the ultra-faint MDF towards lower [Fe/H] values is 
a result of the lower metallicity of the MW environment at their higher formation 
epochs. Indeed UFs form at $z>8.5$ when {\it pre-enrichment} of the MW environment 
was [Fe/H]$<-3.5$ while Sculptor-like dSphs finally assembled at later times, when 
[Fe/H]$\approx -3$ \cite{SFS08}. Note however that the model predicts a small 
[Fe/H]$< -3$ tail in the Sculptor MDF \cite{SF09} that now has been confirmed by 
the new results of the DART team \cite{Else}. What is then the origin of these 
extremely metal-poor stars? 
\section{Merging and star formation histories}                         
\begin{figure}[ht]
  \includegraphics[height=.41\textheight]{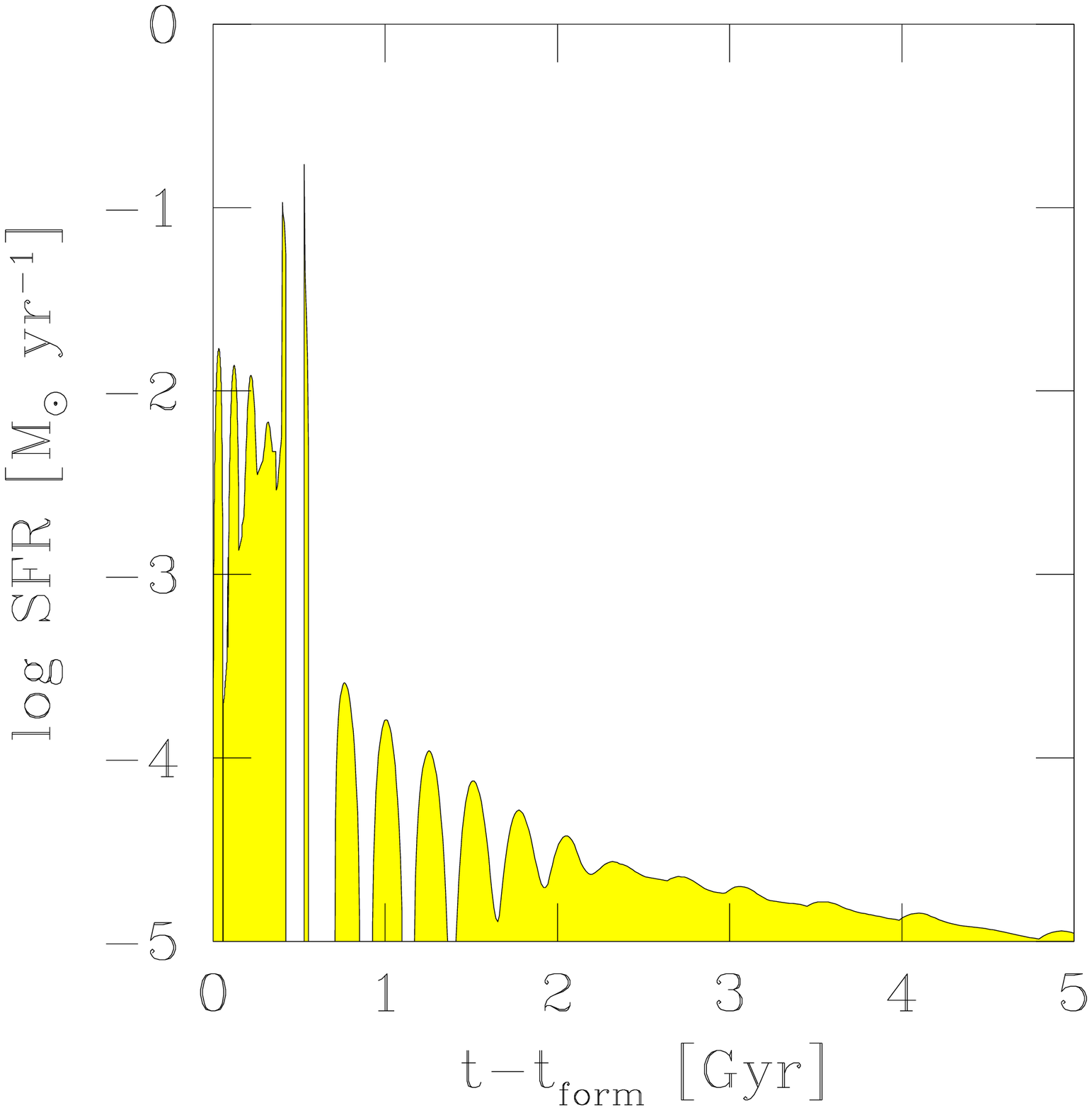}
  \caption{Star formation history of a typical Sculptor-like dSph galaxy.}
  \label{fig:SF}
\end{figure}
We can answer to this question by analyzing the star formation history of a 
typical Sculptor-like dSph (Fig.~5) which exhibits several bursts of different 
intensity and duration. The highest peak corresponds to the SF 
activity at the final assembling epoch ($z\approx 6.5$). Mechanical feedback 
is so efficient at that time (because of the high SF rate) that the galaxy blow 
out the entire gas reservoir in few hundred of Myr, thus suddenly stopping the SF. 
Fresh gas returned by evolved stars is gradually collected after this event, and 
the SF re-start again until a new blow away occurs \cite{SFS08}. However the rates 
are extremely low ($SF<0.0005\Msun$/yr) because of the paucity of returned gas, and 
only a few percent of the total stellar mass is formed during this phase. At earlier 
times ($t-t_{form}<0.5$~Gyr) instead, the peaks reflect the star formation and merging 
activity of the Sculptor progenitor haloes. We find that the oldest progenitors are 
{\it minihaloes} that form stars at very high redshifts $z > 9$ and with typical SF 
rates $<0.02\Msun$/yr. The extremely metal-poor stars that populate the lowest 
metallicity tail of the Sculptor MDF formed in these objects, and hence have {\it 
the same origin of those found in ultra-faint galaxies.}
\section{Conclusions}
Stellar archeology of the most metal-poor stars may provide fundamental insights 
on the early cosmic star formation. In particular the properties of the observed 
Metallicity Distribution Functions in nearby dSph galaxies can be used to constrain 
the origin of these Galactic satellites and to study the physical processes occurring
at high redshifts.
Ultra Faint dSphs are expected to be the today-living counterpart of high redshift
H$_2$-cooling minihaloes ($M=10^{7-8}\Msun$) which are very ineffectively SF objects 
turning into stars $<3\%$ of the potentially available cosmic baryons \cite{SF09}. 
This result is consistent with recent {\it complementary studies} \cite{Bovill,Munoz} 
that use simulations to investigate the origin of dSphs and account for the luminosity 
function of the MW satellites.\\ 
According to our picture the higher fraction of [Fe/H]$<-3$ stars in ultra-faint  
with respect to classical dSphs reflects both the lower SF rate, caused by ineffective 
H$_2$ cooling, both the lower metal (pre-)enrichment of the MW-environment at their 
further formation epoch. Classical dSphs, indeed, are expected to be associated with 
massive H-cooling haloes ($M>10^8\Msun$) that more efficiently cool-down their gas and 
that form at later times ($z<7.5$) through merging of many progenitor haloes. 
Although extremely rare [Fe/H]$ < -3$ stars are expected to populate {\it all} classical 
dSphs. Indeed these stars are expected to mainly form at high redshifts in H$_2$-cooling 
minihaloes, that at later times naturally assemble in more massive dSphs. These findings 
can explain both the detection of [Fe/H]$<-3$ stars in nearby classical dSphs \cite{Martin} 
and the existence of an old stellar population in {\it all} of them \cite{Eline}.

\end{document}